# An autopilot for energy models – automatic generation of renewable supply curves, hourly capacity factors and hourly synthetic electricity demand for arbitrary world regions

Niclas Mattsson, Vilhelm Verendel, Fredrik Hedenus and Lina Reichenberg

## Abstract

Energy system models are increasingly being used to explore scenarios with large shares of variable renewables. This requires input data of high spatial and temporal resolution and places a considerable preprocessing burden on the modeling team. Here we present a new code set with an open source license for automatic generation of input data for large-scale energy system models for arbitrary regions of the world, including sub-national regions, along with an associated generic capacity expansion model of the electricity system. We use ECMWF ERA5 global reanalysis data along with other public geospatial datasets to generate detailed supply curves and hourly capacity factors for solar photovoltaic power, concentrated solar power, onshore and offshore wind power, and existing and future hydropower. Further, we use a machine learning approach to generate synthetic hourly electricity demand series that describe current demand, which we extend to future years using regional SSP scenarios. Finally, our code set automatically generates costs and losses for HVDC interconnections between neighboring regions. The usefulness of our approach is demonstrated by several different case studies based on input data generated by our code. We show that our model runs of a future European electricity system with high share of renewables are in line with results from more detailed models, despite our use of global datasets and synthetic demand.

## 1. Introduction

Models to study energy futures have been developed since the 1970s (Nordhaus, Houthakker, and Solow 1973; Schrattenholzer 1981). The first models had a coarse representation of details of the system, but they have steadily grown in complexity along with advances in computer hardware and improved algorithms in optimizing software. One branch of these models represents the electricity system. The recent realization that variable renewable energy sources may become, or in many cases already are, cheaper than thermal generation has put additional demands on power system models. Increased shares of variable renewables require that power system models represent regional resource endowment, as well as temporal variation of wind and solar output. In response to this development, models with a good representation of wind- and solar variability, as well as other renewable sources, such as hydro power and geothermal power, have been developed (Brown, Hörsch, and Schlachtberger 2018; Pleßmann and Blechinger 2017).

For these models a considerable amount of the work is required to gather and process data, especially weather data. The data input is often tied to the design of the model, or to a specific region. As a model developer, there is thus a choice between doing the data work oneself, which can be a daunting project, or accepting all the assumptions (technical specifications, study area, regionalization, etc.) of another modeler.

Currently there is a growing movement for openness of structure, as well as data input, for energy models (Bazilian et al. 2012; Pfenninger et al. 2017; Weibezahn and Kendziorski 2019; Wiese et al. 2014). Apart from the transparency that allows other researchers to scrutinize assumptions and methods, an additional incentive behind open-source modelling is to enable different groups in society to model and learn from models, and potentially collaborate on the model development itself (Bazilian et al. 2012).

There are a number of open databases for renewable energy, e.g. REAtlas (Andresen, Søndergaard, and Greiner 2015) , Atlite (Github (Atlite) n.d.) and Renewables Ninja (Renewables.ninja n.d.; Pfenninger and Staffell 2016; Staffell and Pfenninger 2016) for wind and solar PV time series and Liu et al. (2019) for global hydropower. Approaches and datasets for producing synthetic electricity demand have also been presented (Toktarova et al. 2019).

REAtlas is a database with wind- and solar data similar to the one presented here. It has global coverage and it is flexible, e.g. in terms of regionalization and turbine functions. In addition, it covers 32 years of weather data. It does not appear to be directly available online, but a version of REAtlas called Atlite is available at Github. Atlite expands on REAtlas in several ways, notably by estimating time series for solar heating, hydropower (run-off) based on Liu et al. (2019), and heating demand. It also includes alternative sources for reanalysis data.

The Renewables Ninja package allows for download of regionally aggregated hourly capacity factors for solar PV and wind power in European countries or their NUTS-2 subregions. Alternatively, hourly capacity factors for an arbitrary point on earth can be downloaded using the web interface. Regionally aggregated capacity factors outside Europe are not currently available. No data is currently provided for concentrated solar power (CSP).

In this paper we aim at making new contributions compared to previous datasets and models, while addressing both the data challenge and the call for open models:

1) **Spatial flexibility**: We generate input data for energy models based on consistent global data sets for wind, solar and CSP for any region of the world. Model regions can be defined using names or identification codes of subregions in the GADM (global) or NUTS2 (Europe) databases of administrative areas, which allows for aggregation levels "from county to country".

2) **Renewable potentials and cost-supply curves**: In addition to hourly capacity factors, we also estimate maximum potential installed capacity for each renewable technology. The potentials are disaggregated by region and resource class, and therefore generate regional cost-supply curves for the downstream energy model.

3) **Techno-economic flexibility**: The costs, turbine function, resource classes, assumptions on land availability for wind and solar are flexible and can easily be changed by the modeler. Included is also representation of current hydro plants and cost-supply curves for further developments.

4) **Synthetic demand** generation based on machine learning. Using input data on temperature and GDP, hourly electricity demand is generated for any region in the world.

In the paper we also demonstrate the model package by presenting

- Validation of the synthetic demand generated
- Cost-supply curves for wind and solar in Europe and China
- Cost and supply mix for a renewable power system in Europe and China.

This paper is structured as follows. In section 2 we give an overview of our GIS and synthetic demand methodology and present our electricity system model. In section 3 we validate the synthetic demand module, construct renewable supply curves and show model results. We discuss our methodology and results in sections 4 and conclude in section 5.

# 2. Method

## 2.1 GIS-based estimates of supply curves and capacity factors

We use a GIS-based approach to estimate parameters required to model renewable energy supply variability on a regional or local level. By "GIS-based", we mean the use of multiple global geospatial datasets combined with global parameter assumptions, without relying on studies of renewable costs and potentials in individual countries. This main advantage of this top-down approach is global consistency of the generated input data. Our modeling framework attempts to reconcile the inherent conflict between large-scale modeling of country- or even continent-sized regions with the high temporal and spatial resolution of solar- and wind supply that is required to adequately model future scenarios with very high penetration of renewables.

The following is a summary of our GlobalEnergyGIS data. For more information, see the online supplementary material.

Capturing regional differences in renewable supply curves and the correlation between temporal profiles of renewable output and electricity demand is of critical importance when modeling renewable based electricity systems. To estimate renewable supply costs and hourly capacity factors in a consistent manner, we perform a GIS-based analysis using ERA5 reanalysis data (Copernicus 2017) combined with auxiliary geospatial datasets: administrative borders (GADM n.d.; Eurostat NUTS n.d.), gridded population (Gao 2017) and GDP (Murakami and Yamagata 2019) in SSP scenarios (Riahi et al. 2017), land cover (Friedl et al. 2010), topography (Amante and Eakins 2009) and protected areas (IUCN 2019).

We use auxiliary datasets to remove pixels which cannot be used for large-scale wind- and solar plants, e.g. due to being in a protected area or having too high population density. The remaining pixels are divided into several resource classes (configurable, but five by default) based on annual capacity factors, and finally we assume a certain fraction of the area of each remaining pixel is available for renewable power. Our baseline assumptions use 5 percent for solar PV and CSP, 5 percent for rooftop PV in urban areas, 8 percent for onshore wind power, and 33 percent for offshore wind power (which is also subject to restrictions on water depth and distance to shore and the nearest electricity grid). Alterative area assumptions can easily be provided by the user. The resulting areas are converted to potentials in GW capacity for each region and resource class using typical power densities (in $W/m^2$) for solar and wind farms.

To estimate potentials for remote solar and wind plants which would require additional transmission investments, we use gridded population (Gao 2017) and purchase-power adjusted gridded GDP (Murakami and Yamagata 2019) to produce a proxy dataset representing grid access, see Supplementary material for details. This proxy is used to further divide our five resource classes for solar PV and onshore wind power into two categories corresponding to local (within 300 km of the grid proxy) or remote (more than 300 km); the latter requires additional investments in transmission.

Finally, the ERA5 reanalysis dataset is used calculate representative time series of aggregated electricity output, i.e. the average capacity factor for each region, resource class and hour. Thus, an investment of, e.g., 1 GW of class 5 onshore wind power in a region will have an hourly output profile corresponding to the average profile of all class 5 wind power pixels in that region. The implicit assumption is that solar and wind plants with similar annual output are distributed evenly throughout each model region. To mitigate bias for wind power, ERA5 hourly time series are scaled to match annual average capacity factors from the Global Wind Atlas (DTU 2019) on a per-pixel basis. This enables us to capture geographical variations in wind power output caused by local differences in topography and land cover at a spatial resolution of 1 km (compared to 31 km for ERA5). For concentrated solar power (CSP), hourly capacity factors for direct solar insolation are produced as above. Our associated energy model optimizes electricity generation from CSP using hourly insolation as input along with other configurable technical parameters, e.g. nine hours of thermal storage and a solar collector multiple of three.

For this analysis we use the new ERA5 reanalysis dataset to generate regional potentials and hourly output profiles for renewables. The ECMWF released the first years of data of ERA5 in July 2017. For this reason, it has not yet been as extensively validated in the literature as its predecessor ERA-Interim. Urraca et al. (2018) use ground observations from 40 BSRN stations to evaluate the accuracy of ERA5 estimates of global horizontal irradiance for solar energy applications. They found that ERA5 provides "a substantial quality leap" over previous datasets. They conclude that ERA5 is now on par with satellite data in inland areas, but that its spatial resolution (31 km) is still too coarse to adequately capture cloud variability in some coastal and mountainous areas.

Olauson (2018) compares ERA5 with MERRA-2 for predicting aggregated wind power generation in five different European countries, and for the output of 1051 individual wind turbines. He finds that ERA5 improves correlations and reduces average errors by 20%, and concludes that ERA5 performs significantly better than MERRA-2 "in all analysed aspects".

For hydropower, we combine public databases on currently existing plants (World Resources Institute 2018) and dams (Lehner et al. 2011) with future potentials, costs, reservoir size and monthly inflow from Gernaat et al. (2017). All three datasets are geospatial with near-global coverage, which enables us to estimate existing capacity and future potential consistently throughout our model regions, including subregions of China. The Gernaat dataset does not extend beyond 60 degrees northern latitude, so the data is supplemented with information on existing hydropower in Scandinavia. The databases for existing hydropower cover approximately 80% of current plants globally. For many countries the coverage is 100%, but others have coverage below 50%. To account for missing plants we scale the individual database entries so the totals match existing national capacity in World Energy Council (2016). Taken together, this approach produces 17 classes of hydro power – one class for existing hydropower, and the remaining 16 from combinations of four cost classes and four reservoir sizes (the smallest for run-of-river plants).

## 2.2 Synthetic demand

Data on hourly electricity demand is currently unavailable for many countries, and even when current time series are available it is not clear how they can be extrapolated into the distant future when GDP and other fundamentals are expected to change. We therefore create future demand scenarios by generating synthetic hourly electricity demand time series (load curves) using a machine learning approach. We take time series of hourly electricity demand for 44 countries from Toktarova et al. (2019) and fit a gradient boosting regression model (Friedman 2001) to demand time series for each country normalized to their annual mean. Estimates of annual country-level annual electricity generation in 2050 were produced by extrapolating annual demand in 2016 (International Energy Agency n.d.) using regional demand growth in the SSP2-34 scenario (Riahi et al. 2017). Our machine learning approach therefore generates predictions of annual and hourly load curves relative to a separately estimated mean.

The profile of electricity demand of a country depends on many variables, such as climate, GDP, industrial structure, technologies used for heating and cooling, tourism etc. Some variables are rather straightforward to predict in the future, whereas others are more difficult. Here we aim to find the profile of the demand based on easily accessible data that can be found for any country, and in future scenarios. Therefore we chose to train our model on seven independent variables: (i) purchase-power adjusted GDP (for prediction, we extrapolated this to 2050 using the SSP2 scenario in a similar way to demand), (ii) average hourly temperature profiles over the year in the 3 most densely populated areas of each country (CIESIN 2016; Gelaro et al. 2017), (iii) the 1st temperature percentile across the year (to represent how low the temperature dips go), (iv) the 99th percentile (to represent how high temperature spikes go), (v) hour of the day, (vi) a weekday/weekend indicator, (vii) a temperature-based ranking of months of the year, where the first month is the coldest month, and the month ranked last is the warmest across the year. The temperature ranking of months was chosen in order to reflect that different countries have summer in different calendar months.

## 2.3 Energy system model and cost data

We have developed a generic electricity system model that directly accepts input from GlobalEnergyGIS, so it can quickly be run for arbitrary region setups. The model conforms to standard practices so the description below is brief. Model equations are listed in the supplementary material.

The model is a capacity expansion model and optimizes investment and dispatch for the electricity sector with hourly time resolution. It employs overnight investment in a greenfield optimization approach. The exception is hydropower, where existing hydropower plants may be assumed to be still in operation. New hydro power is also an investment option in the model, with regional potentials and costs estimated with the GIS methodology outlined above. Technologies are mainly represented by their investment costs, running costs and hourly capacity factors per region and resource class (for wind, solar and hydro power). The model minimizes total system cost subject to demand constraints and a global cap on $CO_2$ emissions (expressed in g $CO_2$ per kWh of electricity demand).

The model has optional ramping constraints and is currently a pure linear programming model, see online supplementary material for technical description. Technology costs and efficiencies are listed in Table 1. Currently batteries are the only pure storage option, but solar CSP and hydro also provide discharge-only storage flexibility.

*Table 1. The technology costs used as input parameters to the modeling.*

| Technology | Investment cost [€/kW] | Fixed cost [€/kW/year] | Variable cost [€/MWh] | Fuel cost [€/MWh fuel] | Efficiency |
|---|---|---|---|---|---|
| Gas GT | 500 | 10 | 1 | 22 | 0.4 |
| Gas CCGT | 800 | 16 | 1 | 22 | 0.6 |
| Coal | 1600 | 48 | 2 | 11 | 0.45 |
| Biogas GT | 500 | 10 | 1 | 37 | 0.4 |
| Biogas CCGT | 800 | 16 | 1 | 37 | 0.6 |
| Nuclear | 5000 | 150 | 3 | 3.2 | 0.4 |
| Wind power | 1200 | 43 | 0 | - | - |
| Off-shore wind | 2300 | 86 | 0 | - | - |
| Solar PV | 600 | 16 | 0 | - | - |
| Solar PV rooftop | 900 | 20 | 0 | - | - |
| Solar CSP | 6000 (incl. 12h thermal storage) | 35 | 0 | - | - |
| HVDC converter pair | 180 | 1.8 | 0 | - | 0.986 |
| HVDC cables (land based) | 0.612 [€/kW/km] | 0.0075 [€/kW/year/km] | 0 | - | 0.016 (losses per 1000 km) |
| Hydro | (site specific variable cost) | - | - | - | - |
| Battery | 150 [€/kWh$_{storage}$] | 1.5 [€/kWh$_{storage}$/year] | 0 | - | 0.9 (round trip) |

Transmission capacity investments and hourly interregional energy flows are determined endogenously in the model and are implemented using a traditional network optimization approach without considering power flow. Estimates of HVDC transmission costs and losses are calculated using parameters from Bogdanov and Breyer (2016) based on distances between population-weighted regional centers, and whether the connection is entirely on land or partially marine, see Table 1. This calculation takes place in the GIS package of our model setup. Matrices representing feasible connections between regions along with their costs and losses are automatically generated by our code. Currently we do not estimate intra-regional costs for transmission and distribution. We simply make a copperplate assumption in our capacity expansion model. However, there is an additional cost for exploiting wind- and solar sites located far from the assumed electricity network, see above.

## 2.4 Code availability

All code and associated documentation has been open-sourced and is available online (Github (GlobalEnergyGIS) n.d.) under a permissive MIT license. The GIS code is written in the Julia programming language and the synthetic demand module in Python. The energy system model is also written in the Julia using the JuMP optimization package and is also available online (Github (Supergrid) n.d.).

# 3. Results

## 3.1 Cross-validation of synthetic demand

In this section we evaluate how well our synthetic demand model reproduces load curve demand patterns on hourly, daily and seasonal time scales. This is done using cross-validation; i.e., for each of our 44 different countries we fit the model exclusively using data for the other 43 countries.

As can be seen in Figure 1, the general shape of the load curves over the year (i.e. seasonal variation) is very well reproduced for most countries. The only country with inaccurate assessment of seasonality is, somewhat surprisingly, France. However, the magnitude of daily variations is significantly over- or underestimated for some countries, e.g. Chile, Iceland, Kenya, Saudi Arabia and Sri Lanka. We speculate that training on data from additional low-income countries could improve the accuracy of predictions of daily variability. In Figure 2 we see that the model is generally quite accurate in capturing hourly and daily variations, as well as weekend demand reduction. The mean-squared prediction error across all countries and hours is 8%.

Taken together, we find that the synthetic demand model generally predicts qualitatively and often quantitatively similar load curves for most countries over hourly, daily and seasonal time scales. It is worth to note that besides month, day and time the model is given only a few variables relating to hourly temperature, GDP, and population distribution. This lends credit to the view that the approach can be generalized across larger geographical regions and into the future.

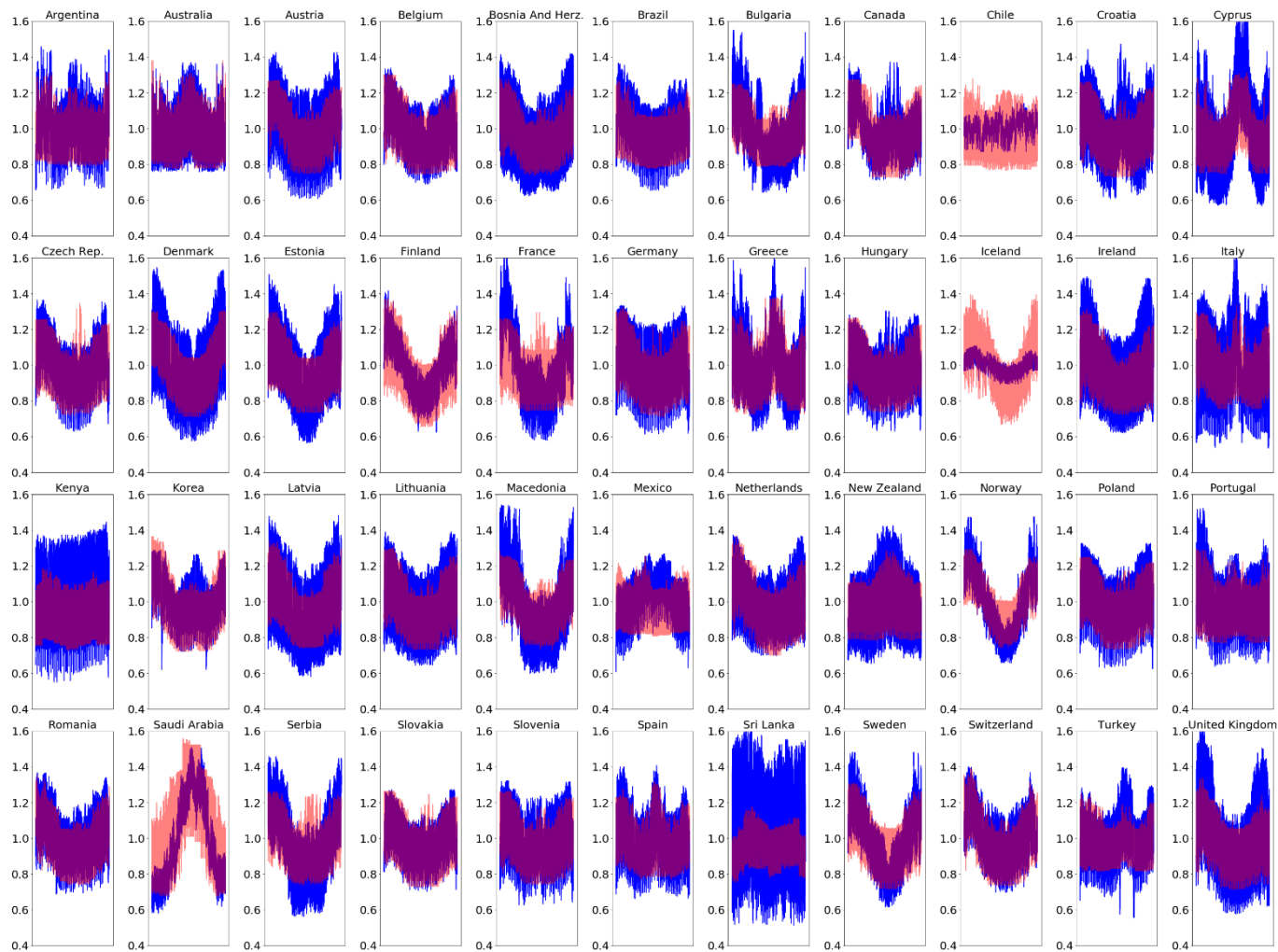

*Figure 1. Cross-validated model predictions (red) compared to real series (blue), normalized with respect to the annual mean. Full year 2015.*

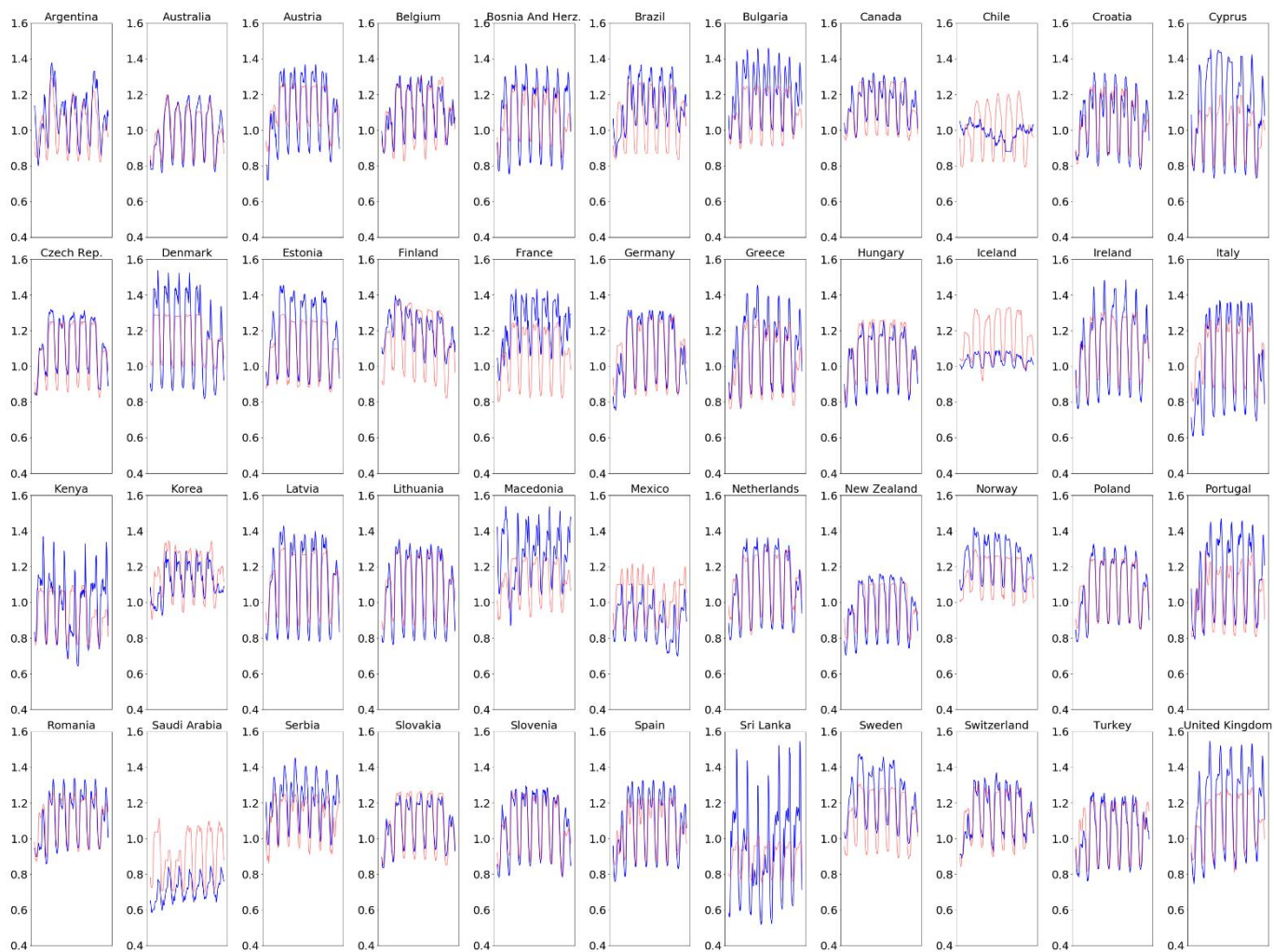

*Figure 2. Cross-validated model predictions (red) compared to real series (blue). One week in January 2015.*

### 3.2 Supply curves for wind- and solar power in Europe

In Figure 3 we present continuous cost-supply curves for solar PV and wind power in Europe and China, constructed by running our GIS code with hundreds of resource classes. The axes show levelized cost of electricity (calculated using technology costs from our energy model, see Table 1) as a function of potential annual electricity generation by solar and wind power, normalized to annual electricity demand in our energy model. Solid lines depict results using our default parameter assumptions and dashed lines show results from doubling the area available for solar- and wind plants (after grid cells were masked out using auxiliary datasets, see section 2.1).

Our results suggest that using default land use assumptions, both solar PV and wind power are in principle capable of supplying 100% of annual electricity demand in Europe at levelized cost below 80 €/MWh. Solar PV is generally available at lower levelized cost than wind, but costs increase more steeply as solar PV approaches 100% supply. This situation is inverted in China, where solar PV can supply over 100% of annual electricity demand at low cost (30-40 €/MWh) but wind power can only supply 80% of annual demand, with costs increasing nearly linearly from 40 to 100 €/MWh. If available land for solar and wind installations is doubled, both solar and wind can individually supply 100% of annual electricity demand in Europe and China at costs below about 60 €/MWh.

The full system cost of a renewables system cannot be assessed based on supply curves alone, as various variation management strategies are required to build a power system fully on solar or wind. Still, the figures provide useful information on the difference in quality of the resources between the two regions, but also the impact of assumptions on how densely solar and wind can be deployed in a region. As shown in Figure 3b, in China, land availability is of little importance for solar supply, but has substantial effect on wind supply.

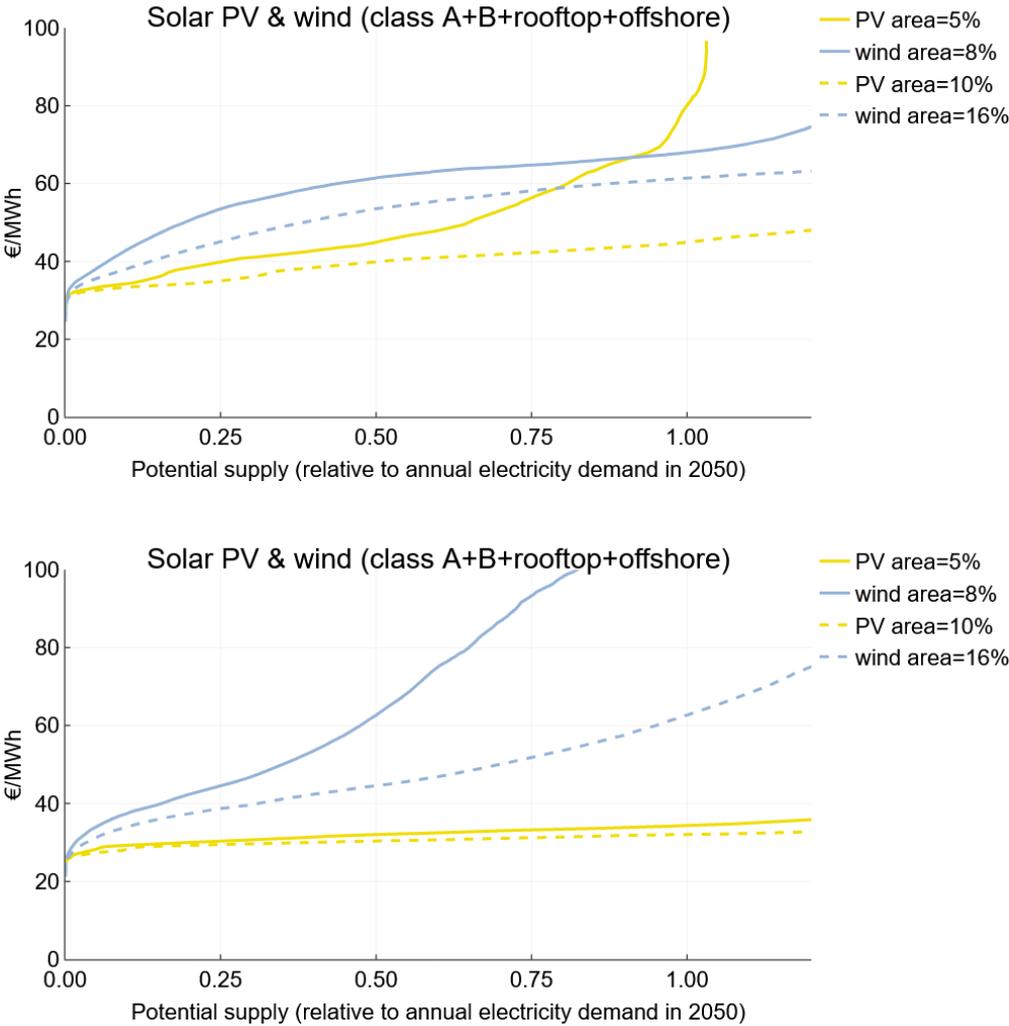

*Figure 3. Cost-supply curves for all solar PV and wind classes in Europe (top) and China (bottom), with default (solid lines) and high assumptions (dashed lines) for land availability. The figure labels show the land area assumptions used for classes A1-A5 (i.e. pixels with electricity access). Land availability in classes B1-B5 (remote areas requiring additional grid investments) is assumed to twice as large.*

## 3.3. Energy mixes in Europe and China with varying land availability

In this section we show results from the full model chain, i.e. using input data generated by the GIS package to run our capacity expansion model of the electricity system. The purpose is to demonstrate that our models produce reasonable results when studying the electricity systems of Europe and China with large shares of variable renewables. More specifically, we analyze how varying assumptions about the renewable resource base and allowed transmission connections affects the optimal energy mix of the system.

First, we run the energy system model using the same assumptions on land availability as shown in Figure 3 in a case with a carbon cap of 25 g $CO_2$ per kWh of annual electricity demand, and with no nuclear or CCS allowed. Resulting energy mixes for Europe and China are shown in Figure 4. We see that high (doubled) land availability generally favors onshore wind power. It primarily replaces offshore wind in Europe, and a mix of solar PV, CSP and offshore wind in China. The preference of onshore over offshore wind when land is relatively scarce is due to the fact that the higher average wind speeds of the offshore resource base are insufficient to compensate for the doubled technology costs compared to onshore wind (Table 1). Figure 4 also demonstrates that wind power has a more advantageous production profile than solar PV, since considerably more wind is installed even though solar PV is available at substantially lower levelized cost than wind power (c.f. Figure 3). This is expected, as solar rather quickly saturates electricity demand during daytime. Additional investments in solar therefore requires storage or other dispatchable capacity which adds to the system cost. For wind power on the other hand, the large interconnected areas in our model (Europe and China) allow for a substantial amount of geographical smoothing, which makes the aggregated production profile flatter and limits the saturation effect.

The main difference between the energy mixes in China and Europe in the default land availability case is the larger deployment of offshore wind in Europe, whereas more CSP is used in China compared to Europe. This is due to differences in regional resource endowment. Europe has large offshore wind potential around the North Sea while CSP is a fairly attractive option in the deserts of western China. However, both of these resources require large amounts of costly long-distance transmission, which is why they are reduced significantly in the high land availability case.

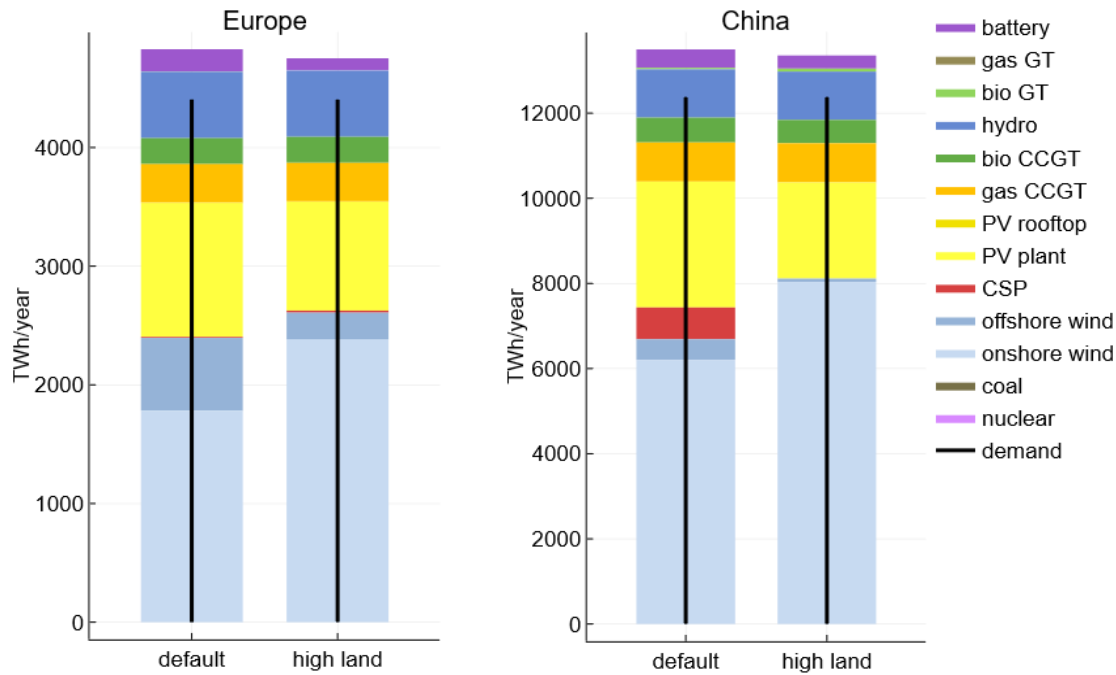

*Figure 4. Annual electricity generation in Europe and China (TWh/year) for default and high land availability cases. Black vertical lines are annual electricity demand. Scenario: carbon cap 25 g $CO_2$/kWh, no nuclear, existing hydro only, unlimited transmission.*

The average electricity cost over all subregions and hours of the year is estimated at 56 €/MWh for Europe in the case of default land availability, and 53 €/MWh for high land availability. Corresponding costs for China are 55 €/MWh in the default case, and 50 €/MWh for the high land availability case.

## 3.4 European system with and without transmission

We also ran an alternative scenario for Europe in which no inter-regional transmission is allowed. The results are compared to the default case with no restrictions on transmission in Figure 5. We note that transmission appears to favor offshore wind power at the expense of CSP. Transmission allows regions around the North Sea to export large amounts of offshore wind power, with installed capacity significantly exceeding local demand. Without transmission, regions that would otherwise import offshore wind power must use more costly local generation, in this case CSP.

Aggregate investments in battery storage in Europe actually decrease somewhat in the case of no transmission. However, the aggregate investments mask opposing changes in individual model regions (see supplementary figure S2). Batteries are generally used as a variation management strategy to shift solar PV power to nighttime hours, and battery investments increase (as expected) in France when solar PV capacity increases. In contrast, in Spain and the Mediterranean region, CSP is favored over the PV + battery combination when no transmission is allowed. In our model CSP includes 12 hours of thermal storage and has no round trip losses, which can partially explain why it is preferred over further increased PV investments in some regions. Indeed, in our model CSP tends to supply more electricity at nighttime than at daytime, see the hourly dispatch in Spain in Figure 6. With thermal storage of CSP, the total amount of storage capacity increases substantially in the no transmission case.

The other reason why CSP is sometimes chosen over PV + batteries may be because of varying probability of clear skies in different regions. In the model, CSP is a substitute to a combination of solar PV and battery storage, except that CSP requires direct solar irradiation while PV also accepts diffuse insolation. Our choice of the ERA5 reanalysis dataset may be significant for this result, since ERA5 distinguishes between direct and diffuse insolation. With other reanalysis datasets that do not make this distinction, direct insolation must be estimated using e.g. the clearness index, which we speculate may be less accurate than if direct and diffuse insolation were separated in the underlying reanalysis model.

Finally, system cost is affected by the availability of transmission. In the scenarios shown in Figure 5, average electricity cost increases from 56 €/MWh (default) to 62 €/MWh (no transmission).

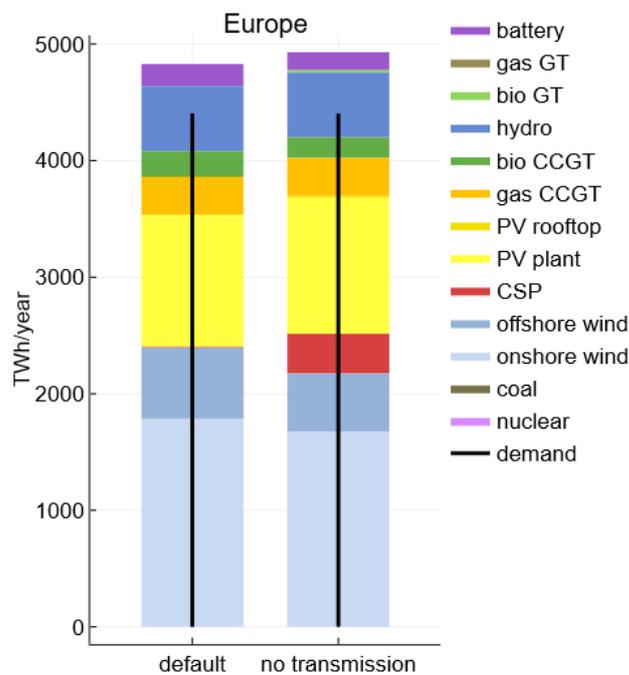

*Figure 5. Annual electricity generation in Europe (TWh/year) for the default case (with unlimited transmission) and an alternative case with no transmission. Black vertical lines are annual electricity demand. Scenario: carbon cap 25 g $CO_2$/kWh, no nuclear, existing hydro only.*

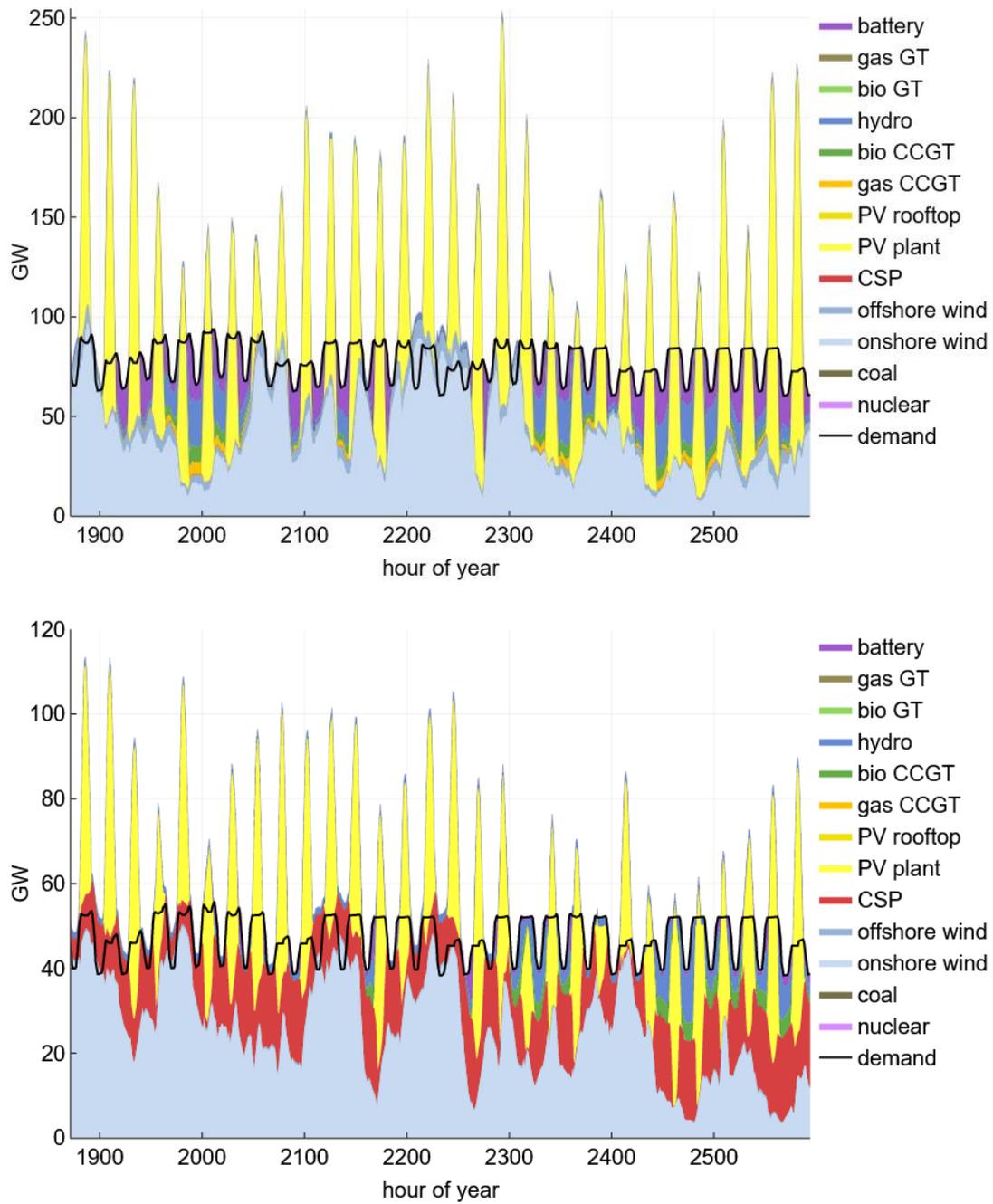

*Figure 6. Hourly electricity dispatch (TWh/hour) during a spring month in France (top) and Spain (bottom). The black line is electricity demand. Scenario: carbon cap 25 g $CO_2$/kWh, no nuclear, existing hydro only, no transmission.*

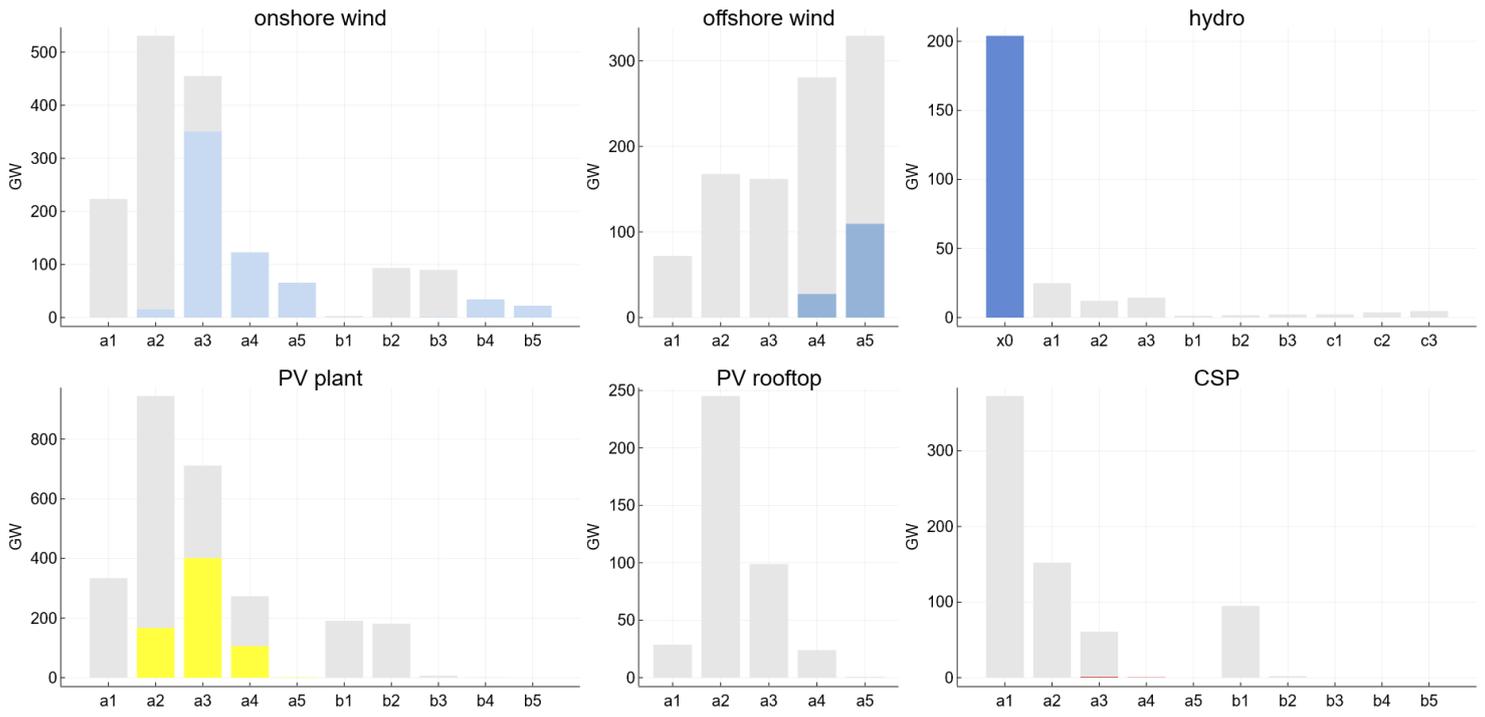

*Figure 7. Realized (colored) and potential capacity (gray) in GW of variable renewable resource classes in Europe (all model regions aggregated). Scenario: carbon cap 25 g $CO_2$/kWh, no nuclear, existing hydro only, ~~no~~ unlimited transmission.*

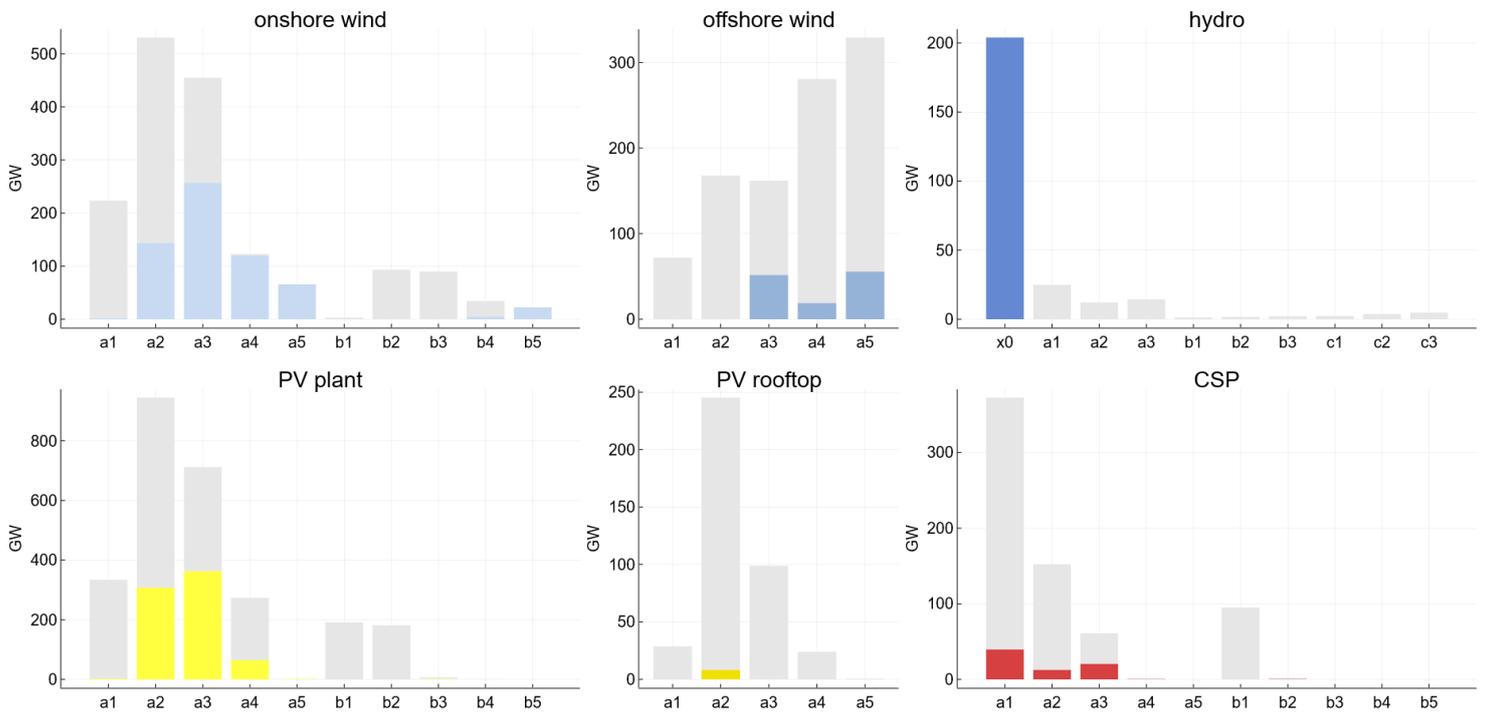

*Figure 8. Realized (colored) and potential capacity (gray) in GW of variable renewable resource classes in Europe (all model regions aggregated). Scenario: carbon cap 25 g $CO_2$/kWh, no nuclear, existing hydro only, no transmission.*

The utilization of resource classes in the default and no transmission scenarios in Europe is shown in Figure 7 and Figure 8. As described in section 2, we divide each resource into five different classes for each model region, and for some technologies we further distinguish between local resources (classes a1-a5, within 300 km of the grid proxy) and remote resources (classes b1-b5, more than 300 km from the grid proxy). As expected, the default case with transmission generally allows more of the high-grade classes to be utilized. However, even in the default case we note some instances in which lower classes are utilized without all the higher-grade resources being fully exploited, see e.g. solar PV and offshore wind. The reason is that the best resources tend to be concentrated in certain regions, and there are costs as well as losses associated with transmitting this electricity to demand centers.

## 4. Discussion

### 4.1 Synthetic demand

As demonstrated by the cross-validation in section 3.1, we generate synthetic load curves for electricity demand with reasonable accuracy based on a limited set of variables for arbitrary countries. However, we trained the algorithm mainly on demand from advanced economies, due to lack of data from developing countries. It is likely that the use of more low-income countries will increase the accuracy of the model (see e.g. Adeoye and Spataru 2019).

Our approach of scaling the demand curve using a projection of annual demand is in line with previous work. Studies have included country variables on different levels of granularity, ranging from industry and sector specific variables describing the uptake of particular technologies such as electric vehicles and heating (Koreneff et al. 2009; Pina, Silva, and Ferrão 2011; Andersen, Larsen, and Boomsma 2013), to models based on more parsimonious time series decomposition (Filik, Gerek, and Kurban 2011). Other models include more detailed country-level data such as public holidays (Sotiropoulos 2012), and model approaches that simulate anomalies (Boßmann and Staffell 2015).

Toktarova et al. (2019) made a similar attempt to estimate the demand profile, partially using the same data as here. However, Toktarova et al. also included variables such as current industrial structure and installed capacity of electrical heating and cooling. Using these features, they were able to reproduce current load patterns with somewhat higher accuracy. However, our goal in this paper is to produce reasonable demand profiles for arbitrary countries extrapolated several decades into the future. We therefore chose to restrict the number and type of variables to a set that can be parameterized using public datasets with global coverage and established SSP scenarios.

### 4.2 Renewable resources

In this study, solar and wind resources were divided into several resource classes for each region, whereas the REAtlas (Andresen, Søndergaard, and Greiner 2015) and Renewables Ninja (Renewables.ninja n.d.) provide one output time series per region. The purpose of resource classes is to give the energy model information about renewable supply curves within each model region by distinguishing between top-, mid- and low-grade resources. This allows the optimization algorithm to exploit the heterogeneity of the resource quality. If only one resource class is constructed per region, the heuristic used to allocate the installations within the area becomes relatively more important compared to if several resource classes are used. On the other hand, if model regions are small, the heterogeneity is also small, and thus the effect of resource classes likely diminishes. However, as an example, many Integrated Assessment Models (IAMs) use model regions of the size of Europe (e.g. Lehtveer, Mattsson, and Hedenus 2017) and Capacity Expansion Models (CEMs) may use regions the size of France (e.g. Bogdanov and Breyer 2016). In such cases, we believe that resource classes are crucial for capturing the priority order and varying potentials of renewable investments within model regions, as well as resolving geographical smoothing of intermittent generation in greater detail.

However, adding resource classes adds computational complexity and thus increases computation time. Each class is roughly equivalent to adding a new technology option in the model, which increases solution time. On the other hand, a model with resource classes can potentially make do with significantly fewer and larger regions, which has the opposite effect on computation time. Exploring the trade-off between resource classes and region size is a research topic in itself. In any case, our GlobalEnergyGIS package can generate input data for models with or without resource classes.

Another explicit feature for renewable resources in the GlobalEnergyGIS package is the set of assumptions regarding land availability for wind- and solar farms. Our results show (Figure 3) that assumptions on land availability influence the supply curve of renewables in a nonlinear fashion and differently for China and Europe. This, in turn, has impact on the system cost of a renewable based power system (Figure 4). Thus, we demonstrate that the model framework is suitable for investigating some of the mechanisms, and related policy issues, which are influential when transforming the power system.

## 4.3 Flexibility in spatial resolution and regional scope

The flexibility in choosing regional sizes and automatic generation of input data makes the model useful for studying a large set of questions and regions. Here we show its potential by comparing China and Europe. The flexibility of the approach has also been demonstrated by the range of case studies it already has been used for, including deep decarbonization scenarios in a future Eurasian supergrid (Reichenberg et al. 2019), benefits of integrating electricity generation in the Middle East and North Africa (Ek-Fälth, Atsmon, and Reichenberg 2019), and a study of nuclear power in the Nordic power system (Kan, Hedenus, and Reichenberg 2019).

The flexibility of geographic scope and granularity also opens up a new set of research questions, where hypotheses formed mainly from studies in a US or European context may be more universally tested. For instance, are transmission extensions crucial to reducing cost for (near) 100% renewable power systems in all parts of the world, as suggested for Europe (Schlachtberger et al. 2017; Brown, Hörsch, and Schlachtberger 2018; Reichenberg et al. 2019), MENA (Ek-Fälth, Atsmon, and Reichenberg 2019), South America (Barbosa et al. 2017) and the USA (MacDonald et al. 2016)? Another question that could be analyzed is the impact of the model artifact introduced by the intra-region copperplate assumption in CEMs. This has been partly explored by Schlachtberger et al. (2017) and Hörsch and Brown (2017), but would merit from a systematic investigation for several continents.

## 4.4 Comparison with other energy system model results for Europe

There are some studies of near 100% renewable power systems in Europe, with which our results for system cost and energy mix may be compared.

The average system cost in this paper with a carbon cap of 25 g $CO_2$/kWh was estimated at 56 €/kWh. Schlachtberger et al. (2017) use a carbon constraint equivalent to 23 g $CO_2$ per kWh and obtain an average electricity cost of 65 €/kWh, whereas Pleßmann and Blechinger (2017) use a more stringent carbon constraint of 5 g $CO_2$/kWh and report an average electricity cost of 88 €/kWh. For a similar set of cost assumptions for wind and solar, Child et al. (2018) found a system cost of 63 €/MWh for a carbon cap of 3 g $CO_2$/kWh. Thus, our approach obtained similar, but slightly lower, system cost than these studies. The reason for the lower cost is probably twofold. First, larger regions reduce the investments required for transmission, which partly explains the difference compared to Schlachtberger et al. (2017) who use country-sized regions. Second, there is a lack of detailed global data on hydropower. We therefore made a somewhat crude assumption on the division (40/60) between run-of-river and reservoir-based hydropower in each region. This assumption may have led to an overestimate of the flexibility of central-European hydropower, which reduces the system cost.

Another way to validate the approach is to analyze the effect of constrained transmission (Figure 5). Not allowing for inter-regional transmission increased the system cost by 10% compared to the case with optimal transmission. A similar estimate was obtained by Pleßmann and Blechinger (2017), in which the system cost was 10% lower for a doubling of transmission capacity compared to the present. Schlachtberger et al. (2017), on the other hand, showed that the difference in system cost between isolated regions and optimal transmission (as in our case) was 27%. The difference between the effect on cost from electrical integration observed by Schlachtberger et al. and the cost difference in the present paper is likely related to the region size: some of the regions in Schlachtberger et al. are rather small, since regions coincide with countries in Europe. Thus, isolating those countries creates model regions in which the renewables resources are scarce, thus increasing costs.

The optimal energy mix in this paper consists of approximately 25% solar and 50% wind for Europe. This energy distribution is similar to Schlachtberger et al. who found a solar/wind penetration of 20% solar and 65% wind.

Overall our simple electricity system model, despite the globally applicable approach, including synthetic electricity demand and less details concerning hydro power, roughly reproduces results from more in-depth studies of Europe using more detailed models. This indicates that the model may be used for other regional analyses in other parts of the world.

# 5 Conclusions

This paper presents GlobalEnergyGIS, a GIS-based package to generate input data for energy system models, and an associated capacity expansion model of electricity supply in arbitrary world regions. To our knowledge there is no model package with the same level of flexibility at the global level as displayed by the effort presented here. Recently there has been a welcome move towards more open energy models and data input (Bogdanov and Breyer 2016; Brown, Hörsch, and Schlachtberger 2018; Huppmann et al. 2019). The transparency and reproducibility that this movement brings are important. The framework presented in this paper aim to further this movement by providing open, consistent data for energy system models at the global level. Our package helps energy modelers:

- easily produce relevant input data on resources and demand for any region of the world
- easily fit spatial granularity (regionalization) and study area depending on research question
- make globally consistent estimates of regional resources that facilitate comparisons between energy futures between different countries
- provide the possibility to analyze any part of the world, including developing countries or other regions which have received less attention from the modelling community.

There are several avenues for further development. In the GIS package, location and infrastructure for fossil fuels and transmission lines, geothermal energy and biomass resources, and carbon storage potential could be added. In the energy system model, a representation of other sectors and technologies would allow for a full sector-coupled model. With these developments an even wider range of research questions could be analyzed.